\documentclass{optica-article}

\journal{opticajournal} % for journals or Optica Open

\articletype{Research Article}

\usepackage{subfigure}
\begin{document}

\title{Probing Axion–Nucleon Coupling with Optomechanical Frequency-Shift Measurements}

\author{Jiawei Li,\authormark{1} Ka-Di Zhu,\authormark{1,*}}

\address{\authormark{1}Key Laboratory of Artificial Structures and Quantum Control (Ministry of Education),
School of Physics and Astronomy, Shanghai Jiao Tong University, 800 DongChuan Road,
Shanghai 200240, China}

\email{\authormark{*}zhukadi@sjtu.edu.cn} %% email address is required; see note below about the corresponding author designation

% use {asbstract*} to suppress the copyright line. Copyright information will be added in production

\begin{abstract*} 
The search for non-baryonic dark matter remains a key focus in modern physics, with the light pseudoscalar axion serving as a well-motivated candidate. Here, we present a laboratory-scale detection scheme to constrain axion-nucleon interactions using a levitated optomechanical sensor, complementing conventional spin-precession and inverse-square-law tests. By monitoring a micro-spherical test mass levitated near alternative aluminum and silver substrate mirrors, our dual-channel differential readout extracts the spin-independent force gradient generated by two-axion exchange. This approach translates the short-range interaction directly into a resolvable splitting in the optical transmission peaks. Our evaluation indicates that for symmetric nucleon coupling ($g_{an}^2=g_{ap}^2$), the dual-cavity platform establishes competitive upper bounds, improving upon existing constraints by up to two orders of magnitude within the $m_{a} \in [0.1, 1]$~eV mass range.
\end{abstract*}

%%%%%%%%%%%%%%%%%%%%%%%%%%  body  %%%%%%%%%%%%%%%%%%%%%%%%%%
\section{Introduction}\label{1}
Extensive astronomical and cosmological observations have indicated the existence of a non-baryonic matter component in the universe, which does not participate in electromagnetic radiation but dominates the gravitational clustering process~\cite{Balbi_2001,chang2022snowmass2021cosmicfrontiercosmic,doi:10.1126/science.284.5419.1481}. The rotational velocities of stars and gas in the outer regions of galaxies cannot be explained solely by the distribution of visible matter. The spatial separation between the gravitational lensing mass peaks and the hot gas distribution in colliding galaxy clusters further indicates that the component dominating the gravitational potential is not identical to ordinary baryons. Furthermore, the precise determination of the cosmic matter composition from the anisotropies of the cosmic microwave background globally supports the cold dark matter (CDM) paradigm~\cite{RubinFord1970,Clowe2006,refId0}. In this context, the axion emerges as a well-motivated class of dark matter candidates. This particle originates from the Peccei--Quinn mechanism proposed to solve the strong CP problem in quantum chromodynamics (QCD), and was subsequently recognized as the light pseudoscalar degree of freedom associated with this spontaneous symmetry breaking~\cite{PecceiQuinn1977,Weinberg1978,Wilczek1978}. Unlike candidates proposed merely out of phenomenological necessity, the mass and coupling strength of the QCD axion are quantitatively related to the symmetry-breaking scale. Moreover, it can yield a relic abundance compatible with cold dark matter via non-thermal production mechanisms in the early universe~\cite{Preskill1983,AbbottSikivie1983}. In the low-energy effective theory, besides coupling to photons, axions can also interact with the axial-vector currents of protons and neutrons. The corresponding axion-nucleon coupling not only determines the form of the additional interactions between nucleons at low-energy scales but also provides a theoretical foundation for a series of precision measurement observables~\cite{diCortona2016,Vonk2020,MoodyWilczek1984}. It should be noted that, for experimental scenarios involving unpolarized macroscopic objects, single-axion exchange typically corresponds to spin-dependent interactions. Conversely, what can be used to describe the additional short-range interactions and perform macroscopic volume integration is usually the spin-independent effective potential induced by two-axion exchange. In this sense, conducting precision measurement studies centered on the axion and its coupling to nucleons provides a concrete pathway for linking microscopic dark matter models to testable experimental signals.

In this work, we propose a high-precision laboratory scheme to constrain spin-independent axion-nucleon interactions using an optically levitated cavity optomechanical platform. By implementing a dual-cavity differential pump-probe configuration, our setup translates the macroscopically integrated force gradient from two-axion exchange into a resolvable mechanical frequency split, while efficiently eliminating common-mode laser noise and environmental perturbations. For symmetric proton and neutron couplings ($g_{an}^{2}=g_{ap}^{2}$), numerical evaluations demonstrate that this scheme yields projected constraints that are improved by up to approximately two orders of magnitude in the axion mass region of $m_{a} \in [0.1~\text{eV}, 1~\text{eV}]$, compared to established limits from spin-precession frequencies~\cite{PhysRevLett.103.261801} and gravitational inverse-square law tests~\cite{PhysRevLett.98.131104,PhysRevLett.98.021101}.

The remainder of this paper is organized as follows. In Sec.~II, we develop the theoretical model, deriving the coordinate-space potential via dispersion relations and performing the volume integration for the sphere-plate configuration. We also establish the Heisenberg equations of motion and the optical transmission coefficient for the dual-cavity layout. Sec.~III provides the detailed estimation of system detection precision, quantifies the Casimir-Polder background along with thermomechanical and momentum-exchange noises, and delineates the resulting projected boundaries in the parameter space. Finally, a concise summary and prospective outlook are outlined in Sec.~IV.

\section{Model and Theory}\label{2}
\subsection{Effective spin-independent potential induced by two-pseudoscalar exchange}\label{2.1}
To investigate the additional interaction mediated by unpolarized objects, we consider the non-derivative Yukawa coupling between a pseudoscalar field $\phi$ and a fermion $\psi_a$:
\begin{equation}
\mathcal{L}_{\mathrm{int}}
=
i g_a \bar{\psi}_a \gamma_5 \psi_a \phi ,
\end{equation}
where $a$ labels the different species of fermions, such as protons, neutrons, or electrons. Performing a Dyson--Foldy transformation on the interaction and retaining the lowest-order terms in the heavy-fermion limit yields the effective interaction Hamiltonian~\cite{PhysRevD.59.075009,PhysRev.91.1527}:
\begin{equation}
H_{\mathrm{eff}}^{\mathrm{NR}}
=
-\frac{g_a}{2M_a}\,
\psi_a^\dagger \bigl(\boldsymbol{\sigma}\cdot\nabla\phi\bigr)\psi_a
+\frac{g_a^2}{2M_a}\,
\psi_a^\dagger \psi_a\,\phi^2
+\cdots ,
\label{eq:Heff_pair}
\end{equation}
where the second term is the scalar pair term that introduces a spin-independent central potential. For brevity, we introduce the notation $G_a \equiv \frac{g_a^2}{2M_a}$ and $G_b \equiv \frac{g_b^2}{2M_b}$. Consequently, the two-exchange amplitude dominated by this term between fermions $a$ and $b$ can be written as~\cite{10.1119/1.17586}:
\begin{equation}
i\mathcal{M}(q)
=
-2i\,G_a G_b\,\Gamma(q^2),
\label{eq:Mq_def}
\end{equation}
where
\begin{equation}
\Gamma(q^2)
=
\int \frac{d^4k}{(2\pi)^4}
\frac{i}{k^2-m^2+i\epsilon}
\frac{i}{(q-k)^2-m^2+i\epsilon},
\label{eq:loop_gamma}
\end{equation}
with $m$ being the pseudoscalar mass and $q$ denoting the four-momentum transfer.

To obtain the coordinate-space potential from the momentum-space amplitude, we employ a dispersion relation. For a central potential that depends solely on the interparticle distance $r$, we have:
\begin{equation}
V(r)
=
-\frac{i}{8\pi^2 r}
\int_{4m^2}^{\infty}
dt\,
\bigl[\mathcal{M}(t)\bigr]\,
e^{-\sqrt{t}\,r},
\label{eq:dispersion_general}
\end{equation}
where $t=q^2$, and $\bigl[\mathcal{M}(t)\bigr]$ represents the discontinuity of the amplitude across the two-particle threshold on the real axis. Since the intermediate state comprises two pseudoscalars each of mass $m$, the threshold is located at $t=4m^2$.

Using the Cutkosky rules~\cite{PhysRevD.49.4951}, the two internal propagators can be cut simultaneously, yielding:
\begin{equation}
\bigl[\Gamma(t)\bigr]
=
\frac{1}{(2\pi)^2}
\int d^4k\,
\delta(k^2-m^2)\,
\delta[(q-k)^2-m^2]\,
\theta(k^0)\,
\theta(q^0-k^0).
\end{equation}
This integral corresponds to the standard two-body phase-space integration, which results in:
\begin{equation}
\bigl[\Gamma(t)\bigr]
=
\frac{1}{8\pi}
\sqrt{1-\frac{4m^2}{t}}.
\label{eq:disc_gamma}
\end{equation}
From Eq.~\eqref{eq:Mq_def}, it follows that:
\begin{equation}
\bigl[\mathcal{M}(t)\bigr]
=
-2i\,G_a G_b\,\bigl[\Gamma(t)\bigr]
=
-\frac{iG_aG_b}{4\pi}
\sqrt{1-\frac{4m^2}{t}}.
\label{eq:disc_M}
\end{equation}
Substituting this back into Eq.~\eqref{eq:dispersion_general} gives:
\begin{equation}
V(r)
=
-\frac{G_aG_b}{32\pi^3 r}
\int_{4m^2}^{\infty}
dt\,
e^{-\sqrt{t}\,r}
\sqrt{1-\frac{4m^2}{t}}.
\label{eq:V_before_int}
\end{equation}

Next, we perform a change of variables for the integration. Setting
\begin{equation}
t=s^2,
\qquad
dt=2s\,ds,
\end{equation}
we obtain:
\begin{equation}
\int_{4m^2}^{\infty}
dt\,
e^{-\sqrt{t}\,r}
\sqrt{1-\frac{4m^2}{t}}
=
2\int_{2m}^{\infty}
ds\,e^{-rs}\sqrt{s^2-4m^2}.
\label{eq:int_s}
\end{equation}
Further letting
\begin{equation}
s=2m\cosh u,
\qquad
ds=2m\sinh u\,du,
\qquad
\sqrt{s^2-4m^2}=2m\sinh u,
\end{equation}
the above expression transforms into:
\begin{equation}
2\int_{2m}^{\infty}
ds\,e^{-rs}\sqrt{s^2-4m^2}
=
8m^2\int_0^{\infty}
du\,e^{-2mr\cosh u}\sinh^2u.
\end{equation}
Utilizing the identity
\begin{equation}
\sinh^2u=\frac{\cosh 2u-1}{2},
\end{equation}
and the integral representation of the modified Bessel function
\begin{equation}
K_\nu(z)=\int_0^\infty du\,e^{-z\cosh u}\cosh(\nu u),
\end{equation}
we find:
\begin{equation}
\int_{4m^2}^{\infty}
dt\,
e^{-\sqrt{t}\,r}
\sqrt{1-\frac{4m^2}{t}}
=
4m^2\bigl[K_2(2mr)-K_0(2mr)\bigr].
\end{equation}
By applying the recurrence relation
\begin{equation}
K_2(z)-K_0(z)=\frac{2}{z}K_1(z),
\end{equation}
we arrive at:
\begin{equation}
\int_{4m^2}^{\infty}
dt\,
e^{-\sqrt{t}\,r}
\sqrt{1-\frac{4m^2}{t}}
=
\frac{4m}{r}K_1(2mr).
\label{eq:int_final}
\end{equation}
Consequently, Eq.~\eqref{eq:V_before_int} finally reduces to:
\begin{equation}
V_{ab}^{(4)}(r)
=
-\frac{G_aG_b\,m}{8\pi^3 r^2}K_1(2mr).
\end{equation}
Substituting the definitions of $G_a$ and $G_b$ back into the expression yields:
\begin{equation}
V_{ab}^{(4)}(r)
=
-\frac{g_a^2g_b^2}{32\pi^3 M_aM_b}
\left(\frac{m}{r^2}\right)K_1(2mr).
\label{eq:two_pseudo_final}
\end{equation}
This is the spin-independent effective potential induced by the two-pseudoscalar exchange.

To evaluate the observable signals in a sphere-plate configuration from the aforementioned effective potential generated by two-axion exchange, it is necessary to sum over all nucleon pairs within the two macroscopic bodies~\cite{PhysRevD.89.035010}. Under the assumption that the additional interaction is significantly weaker than the internal electromagnetic binding energy of the materials, the additivity approximation can be adopted, allowing the total interaction energy to be expressed in a volume integral form. Note that because we compute the signal difference between two scenarios, the influence of the gold-coated layer on each cavity mirror surface cancels out as a uniform background and is thus omitted here. Concurrently, the lateral dimensions of the cavity mirrors are assumed to be sufficiently large compared to the radius of the sphere so that they can be treated as infinite plates. In cylindrical coordinates, the interaction potential experienced by a volume element $dV$ of a $\text{SiO}_2$ nanosphere due to a plate of thickness $D$ and radius $R_p$ is given by:
\begin{equation}
dU(z)=
-\frac{m_a\,C_s C_p}{2\pi m_\text{ave}^2 m_H^2}\int_{z}^{z+D} dz_1 \int_{0}^{R_p}\rho\, d\rho\,
\frac{K_{1}\!\left(2m_a\sqrt{\rho^{2}+z_1^{2}}\right)}
{\rho^{2}+z_1^{2}},
\label{eq:Uz_start}
\end{equation}
where $m_a$ is the axion mass, $z$ is the perpendicular distance from the volume element to the plate surface, and $\rho$ is the radial variable in cylindrical coordinates. For homogeneous materials, the effective coefficients are defined as:
\begin{equation}
C_i
=
\rho_i
\left(
\frac{g_{ap}^2}{4\pi}\frac{Z_i}{\mu_i}
+
\frac{g_{an}^2}{4\pi}\frac{N_i}{\mu_i}
\right),
\qquad i=s,p,
\end{equation}
where the subscripts $s$ and $p$ denote the sphere and the plate, respectively; $\rho_i$ represents the material density, $Z_i/\mu_i$ and $N_i/\mu_i$ characterize the relative abundance of protons and neutrons per unit mass, $\mu_i=m_i/m_H$, with $m_H$ being the mass of a hydrogen atom; and $m_\text{ave}=(m_p+m_n)/2$ is the average nucleon mass. In this manner, under the assumption of bulk homogeneity, the contribution from each part of the plate is accounted for via the nucleon number density.

Differentiating Eq.~\eqref{eq:Uz_start} with respect to $z$ yields the additional force acting on a single volume element $dV$:
\begin{equation}
dF_{\mathrm{add}}(z)
=
-\frac{\partial U(z)}{\partial z}.
\label{eq:dFda_def}
\end{equation}
Applying Leibniz's rule for differentiation under the integral sign,
\begin{equation}
\frac{d}{dz}\int_{z}^{z+D}f(z_1)\,dz_1=f(z+D)-f(z),
\end{equation}
we obtain:
\begin{align}
dF_{\mathrm{add}}(z)
&=
-\frac{m_a\,C_s C_p}{2\pi m_\text{ave}^2 m_H^2}
\int_{0}^{R_p}\rho\,d\rho
\Biggl[
\frac{K_{1}\!\left(2m_a\sqrt{\rho^{2}+z^{2}}\right)}{\rho^{2}+z^{2}}-\frac{K_{1}\!\left(2m_a\sqrt{\rho^{2}+(z+D)^{2}}\right)}{\rho^{2}+(z+D)^{2}}
\Biggr].
\label{eq:Fadd_rho}
\end{align}

Next, we adopt the integral representation of $K_1$:
\begin{equation}
K_1(t)=t\int_{1}^{\infty} e^{-tu}\sqrt{u^{2}-1}\,du.
\label{eq:K1_repr}
\end{equation}
For the first term in the integrand of Eq.~\eqref{eq:Fadd_rho}, we define the new variables:
\begin{equation}
t=2m_a\sqrt{\rho^{2}+z^{2}},
\qquad
t_{0}^{(1)}=2m_a z,
\qquad
t_{R_p}^{(1)}=2m_a\sqrt{R_p^{2}+z^{2}},
\end{equation}

Then, from 
\[
t=2m_a s,\qquad s=\sqrt{\rho^{2}+z^{2}},
\qquad
\rho\,d\rho=s\,ds=\frac{t\,dt}{4m_a^{2}},
\]
we obtain
\begin{align}
\int_{0}^{R_p}\rho\,d\rho\,
\frac{K_1\!\left(2m_a\sqrt{\rho^{2}+z^{2}}\right)}{\rho^{2}+z^{2}}
&=
\int_{0}^{R_p}\rho\,d\rho\,
\frac{K_1(t)}{s^{2}}
\nonumber\\
&=
\int_{t_{0}^{(1)}}^{t_{R_p}^{(1)}} \frac{K_1(t)}{t}\,dt
\nonumber\\
&=
\int_{t_{0}^{(1)}}^{t_{R_p}^{(1)}}dt
\int_{1}^{\infty} e^{-tu}\sqrt{u^{2}-1}\,du
\nonumber\\
&=
\int_{1}^{\infty}\sqrt{u^{2}-1}\,du
\int_{t_{0}^{(1)}}^{t_{R_p}^{(1)}} e^{-tu}\,dt.
\end{align}
Similarly, by defining the variables for the second term in the integrand of Eq.~\eqref{eq:Fadd_rho} as
\begin{equation}
t=2m_a\sqrt{\rho^{2}+(z+D)^{2}},
\qquad
t_{0}^{(2)}=2m_a(z+D),
\qquad
t_{R_p}^{(2)}=2m_a\sqrt{R_p^{2}+(z+D)^{2}},
\end{equation}
one finds
\begin{equation}
\int_{0}^{R_p}\rho\,d\rho\,
\frac{K_{1}\!\left(2m_a\sqrt{\rho^{2}+(z+D)^{2}}\right)}
{\rho^{2}+(z+D)^{2}}
=
\int_{1}^{\infty}\sqrt{u^{2}-1}\,du
\int_{t_{0}^{(2)}}^{t_{R_p}^{(2)}} e^{-tu}\,dt.
\end{equation}

Consequently, the additional force can be rewritten as
\begin{align}
dF_{\mathrm{add}}(z)
&=
-\frac{m_a\,C_s C_p}{2\pi m_\text{ave}^2 m_H^2}
\int_{1}^{\infty}\sqrt{u^{2}-1}\,du
\left[
\int_{t_{0}^{(1)}}^{t_{R_p}^{(1)}} e^{-tu}\,dt
-
\int_{t_{0}^{(2)}}^{t_{R_p}^{(2)}} e^{-tu}\,dt
\right].
\end{align}
Utilizing the identity $\int e^{-tu}\,dt=-\frac{1}{u}e^{-tu}$, we yield
\begin{align}
dF_{\mathrm{add}}(z)
&=
-\frac{m_a\,C_s C_p}{2\pi m_\text{ave}^2 m_H^2}
\int_{1}^{\infty}\frac{\sqrt{u^{2}-1}}{u}\,du
\Bigl[
e^{-t_{0}^{(1)}u}
-
e^{-t_{R_p}^{(1)}u}
-
e^{-t_{0}^{(2)}u}
+
e^{-t_{R_p}^{(2)}u}
\Bigr].
\label{eq:Fadd_u}
\end{align}

Considering that the dimensions of the plate (i.e., the cavity mirror in the experimental system) are much larger than the other geometric parameters considered here, the terms $e^{-t_{R_p}^{(1)}u}$ and $e^{-t_{R_p}^{(2)}u}$ can be safely neglected. Integrating over the entire volume of the nanosphere yields the final expression:
\begin{equation}
U(z)
=
-\frac{C_s C_p m_a}{2m_\text{ave}^2 m_H^2}
\int_1^\infty du\,
\frac{\sqrt{u^2-1}}{u}
\left(1-e^{-2m_auD}\right)
e^{-2m_a u z}
\int_0^{2r_s} dz\,A(z),
\label{eq:Ua_slice}
\end{equation}
where $A(z)=\pi(2r_s z-z^2)$ represents the cross-sectional area of the sphere with radius $r_s$ at a height $z$ from its bottom.

Substituting Eq.~\eqref{eq:Ua_slice} into Eq.~\eqref{eq:dFda_def} leads to the force gradient:
\begin{equation}
\frac{\partial F_\text{add}(z)}{\partial z}
=
\frac{2\pi r_s^3 m_a}{3m_\text{ave}^2 m_H^2}\,C_s C_p
\int_1^\infty du\,
\frac{\sqrt{u^2-1}}{u^2}
\left(1-e^{-2m_auD}\right)
e^{-2m_auz}.
\label{eq:dFda_general}
\end{equation}

Equation~\eqref{eq:dFda_general} demonstrates that the force gradient can be naturally factored into two parts: a material composition factor $C_s C_p$ and an exponential kernel governed by the interaction range and the plate thickness. This completes the derivation connecting the effective inter-nucleon potential induced by two-axion exchange to the analytical expression for the additional force gradient in a sphere-plate configuration. The core of the subsequent numerical calculation lies precisely in the evaluation of Eq.~\eqref{eq:dFda_general} under various coupling hypotheses.
\subsection{A scheme for detecting force gradient anomalies based on a dual-cavity mirror system}\label{2.2}
\begin{figure}[!h]
\centering
\includegraphics[width=1\textwidth]{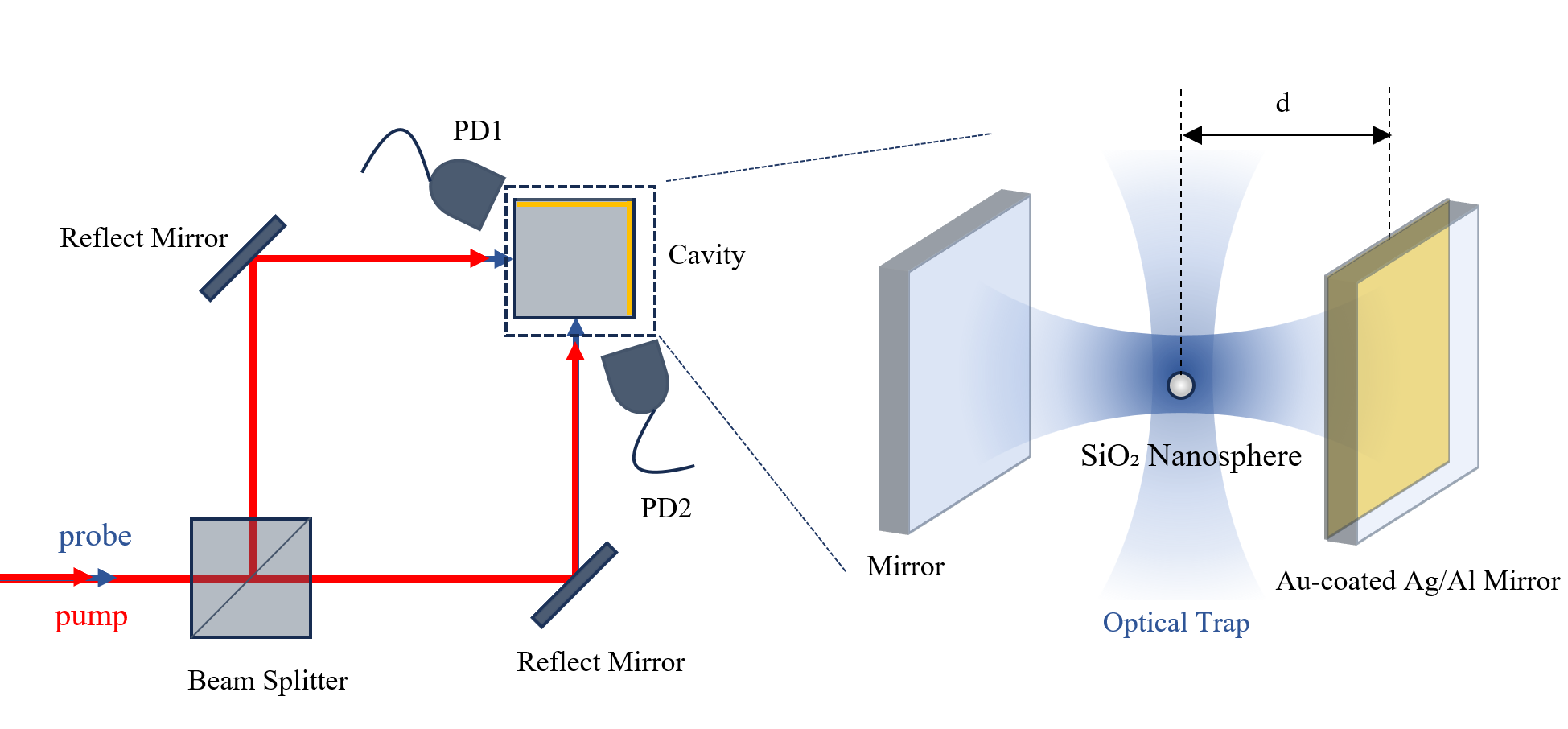}
\caption{Schematic diagram of the experimental setup. Left: Optical layout based on the pump--probe scheme. A laser beam containing both pump and probe components is incident on a 50:50 beam splitter, where it is divided into two mutually orthogonal beams. These two beams are directed into two orthogonal cavity axes of a vacuum chamber, each employing mirrors with different substrates: one is a gold-coated aluminum substrate mirror, and the other is a gold-coated silver substrate mirror. The external photodetection readout system is omitted here for clarity. Right: Structure of a single cavity axis. Within an optical cavity formed by two mirrors separated by a distance $d$, a SiO$_2$ nanosphere is optically levitated using optical tweezers.}
\label{fig:5.1}
\end{figure}
Here, we consider the levitated cavity optomechanical system illustrated in Fig.~\ref{fig:5.1}, where the pump and probe beams enter two independent optical cavities after passing through a beam splitter. Performing two consecutive experiments using a single cavity is highly susceptible to temporal drifts and non-repeatable errors. In contrast, simultaneously injecting the same laser beam into two distinct cavities via a beam splitter converts time-dependent perturbations—such as laser frequency drift, power fluctuations, and environmental variations—into a common-mode background shared by both channels, which can be effectively canceled out through differential comparison. The two optical cavities, configured with identical geometric and experimental parameters, each consist of a standard mirror and a gold-coated aluminum or silver substrate mirror that serves as the source mass. The Hamiltonian of this system can be expressed as~\cite{PhysRevA.77.033804,PhysRevA.63.023812}:
\begin{align}
H ={} & \hbar \omega_m b^\dagger b + \hbar \omega_c c^\dagger c + \hbar g (b^\dagger + b)c^\dagger c + i\hbar \Omega_\text{pu}( c^\dagger e^{-i\omega_\text{pu} t} - c e^{i\omega_\text{pu} t}) \nonumber \\
& + i\hbar \Omega_\text{pr}( c^\dagger e^{-i\omega_\text{pr} t} - c e^{i\omega_\text{pr} t}),
\end{align}
where $\omega_{m}$ is the resonance frequency of the nanosphere mechanical resonator, with $b^\dagger$ ($b$) being the corresponding creation (annihilation) operator; $\omega_{c}$ is the resonance frequency of the cavity mode, with $c^\dagger$ ($c$) being the corresponding creation (annihilation) operator; $g$ characterizes the optomechanical coupling strength between the cavity field and the nanosphere; $\omega_\text{pu}$ and $\omega_\text{pr}$ are the frequencies of the pump and probe lasers, respectively; and the Rabi frequencies $\Omega_\text{pu}$ and $\Omega_\text{pr}$ are related to the laser power $P$ via
$\Omega_\text{pu} = \sqrt{2P_\text{pu}\kappa/\hbar\omega_\text{pu}}$ and 
$\Omega_\text{pr} = \sqrt{2P_\text{pr}\kappa/\hbar\omega_\text{pr}}$,
with $\kappa$ denoting the amplitude decay rate of the cavity field.

In the rotating frame at the driving field frequency $\omega_\text{pu}$, the Hamiltonian transforms into:
\begin{align}
\tilde{H} ={} & \hbar \omega_m b^\dagger b + \hbar \Delta c^\dagger c + \hbar g (b^\dagger + b)c^\dagger c \nonumber\\
& + i\hbar \Omega_\text{pu}( c^\dagger - c) + i\hbar \Omega_\text{pr}( c^\dagger e^{-i\delta t} - c e^{i\delta t}) ,
\end{align}
where $\delta = \omega_\text{pr} - \omega_\text{pu}$ is the pump-probe detuning, and $\Delta = \omega_{c} - \omega_\text{pu}$ is the pump-cavity detuning. Defining the position-like operator $\tau = b + b^\dagger$ and applying the Heisenberg equations of motion, we obtain:
\begin{equation}
\frac{dc}{dt} = -i\Delta c - ig(b^\dagger + b)c + \Omega_\text{pu} + \Omega_\text{pr} e^{-i\delta t},
\end{equation}
and
\begin{equation}
\frac{d^2 \tau}{dt^2} + \omega_m^2 \tau = -2g \omega_m c^\dagger c.
\end{equation}

By incorporating the respective damping terms, Eqs. (3) and (4) can be rewritten as:
\begin{equation}
\frac{dc}{dt} + (i\Delta + \kappa) c = -ig(b^\dagger + b)c + \Omega_\text{pu} + \Omega_\text{pr} e^{-i\delta t},
\end{equation}
and
\begin{equation}
\frac{d^2 \tau}{dt^2} + \gamma_m \frac{d \tau}{dt} + \omega_m^2 \tau = -2g \omega_m c^\dagger c,
\end{equation}
where $\gamma_m$ represents the damping rate of the mechanical resonator.

Taking the expectation values of Eqs. (5) and (6) yields:
\begin{equation}
\left\langle \frac{dc}{dt} \right\rangle + (i\Delta + \kappa) \langle c\rangle = -2ig\langle \tau c \rangle + \Omega_\text{pu} + \Omega_\text{pr} e^{-i\delta t},
\end{equation}
and
\begin{equation}
\frac{d^2 \langle \tau \rangle}{dt^2} + \gamma_m \frac{d \langle \tau \rangle}{dt} + \omega_m^2 \langle \tau \rangle = -2g \omega_m \langle c^\dagger c\rangle.
\end{equation}

We make the following ansatz~\cite{10.5555/1817101}:
\begin{equation}
\langle c(t) \rangle = c_0 + c_+ e^{-i\delta t} + c_- e^{i\delta t},
\end{equation}
\begin{equation}
\langle \tau(t) \rangle = \tau_0 + \tau_+ e^{-i\delta t} + \tau_- e^{i\delta t}.
\end{equation}
Concurrently, the factorization approximations are assumed:
\begin{equation}
\langle c^\dagger c \rangle = \langle c^\dagger\rangle \langle c\rangle , 
\end{equation}
and
\begin{equation}
\langle \tau c \rangle = \langle \tau \rangle \langle c\rangle . 
\end{equation}

Substituting Eqs. (9)--(12) into Eqs. (7) and (8), a series of algebraic evaluations yields:
\begin{equation}
|\Omega_\text{pu}|^2 = \left[ \kappa^2 + \left( \Delta - \frac{2g^2 \sigma}{\omega_m} \right)^2 \right] \sigma,
\end{equation}
where $\sigma$ is defined as $\sigma \equiv |c_0|^2$, and
\begin{equation}
c_+ = \frac{\Omega_\text{pr} O_1 (O_1 O_2 - iO_4)}{(O_1 O_3 - iO_4)(O_1 O_2 - iO_4) + O_4^2},
\end{equation}
with
\begin{align}
O_1 ={} & \omega_m^2 - i\delta\gamma_m - \delta^2,\nonumber\\
O_2 ={} & -\kappa + i\delta + i\Delta - \frac{2ig^2\sigma}{\omega_m},\nonumber\\
O_3 ={} & \kappa - i\delta + i\Delta - \frac{2ig^2\sigma}{\omega_m},\nonumber\\
O_4 ={} & 2g^2\sigma\omega_m.
\end{align}

To investigate the optical properties of the output fields, we utilize the standard input--output relation for a single-ended cavity, $c_\text{out}(t) = c_\text{in}(t) - \sqrt{2\kappa}c(t)$~\cite{bowen2015quantum}, where $c_\text{in}$ and $c_\text{out}$ denote the input and output operators, respectively. This leads to the expectation value of the output field:
\begin{align}
\langle c_\text{out}(t)\rangle ={} & (\Omega_\text{pu}/\sqrt{2\kappa} - \sqrt{2\kappa} c_0)e^{-i\omega_\text{pu}t} \nonumber\\
& + (\Omega_\text{pu}/\sqrt{2\kappa} - \sqrt{2\kappa} c_+)e^{-i(\omega_\text{pu}+\delta)t} \nonumber\\
& - \sqrt{2\kappa}c_-e^{-i(\omega_\text{pu}-\delta)t}.
\end{align}
The transmission coefficient of the probe beam, defined as the ratio of the output field amplitude to the input field amplitude at the probe frequency, is expressed as:
\begin{equation} 
t = \frac{\Omega_\text{pr}/\sqrt{2\kappa} - \sqrt{2\kappa}c_+}{\Omega_\text{pr} /\sqrt{2\kappa}} = 1 - 2\kappa c_+/\Omega_\text{pr}.
\end{equation}

\section{NUMERICAL RESULTS}\label{3}
\subsection{System detection precision estimation}\label{3.1}
This section details the specific set of experimental parameters adopted in the numerical simulations and describes the computational configuration for the cavity transmission spectrum. For the optical system, we consider a plano-concave Fabry--P\'erot optical cavity. The wavelength of the incident laser beam is chosen as $\lambda = 1064\,\mathrm{nm}$, the cavity length is $L = 1\,\mathrm{mm}$, the radius of curvature of the concave mirror is $R_\text{mi} = 40\,\mathrm{mm}$, and the cavity finesse is set to $F = 40$. Consequently, the amplitude decay rate of the cavity field can be determined by~\cite{article3}:
\begin{equation}
\kappa = \frac{\pi c}{2FL} = 2\pi \times 1.87 \times 10^{9}\,\mathrm{Hz},
\end{equation}
and the resonance frequency of the cavity mode is given by:
\begin{equation}
\omega_c = \frac{2\pi c}{\lambda} = 1.77 \times 10^{15}\,\mathrm{Hz}.
\end{equation}
For the plano-concave cavity configuration specified above, the waist radius of the cavity mode can be expressed as:
\begin{equation}
w_c = \sqrt{\frac{\lambda}{\pi}}\,[L(R_\text{mi}-L)]^{1/4}.
\end{equation}
Substituting the parameter values yields $w_c = 4.60 \times 10^{-5}\,\mathrm{m}$, corresponding to a cavity mode volume of:
\begin{equation}
V_c = \frac{\pi L w_c^2}{4} = 1.66 \times 10^{-12}\,\mathrm{m}^3.
\end{equation}

Regarding the mechanical component, we consider a silica nanosphere with a radius of $r_s = 20\,\mathrm{nm}$, a material density of $\rho = 2200\,\mathrm{kg/m^3}$, and a relative permittivity of $\epsilon_r = 3.75$. The volume and mass of the nanosphere are evaluated respectively as:
\begin{equation}
V_s = \frac{4\pi r_s^3}{3} = 3.35 \times 10^{-23}\,\mathrm{m}^3,
\qquad
m = \rho V_s = 7.37 \times 10^{-20}\,\mathrm{kg}.
\end{equation}
The trapping potential is provided by optical tweezers with a trapping laser power of $P_{\mathrm{trap}} = 0.17\,\mathrm{W}$ and a beam waist of $w_{\mathrm{trap}} = 0.9\,\mu\mathrm{m}$ (where $w_{\mathrm{trap}}$ corresponds to $w_0$ in the focal plane). In the experimental design described in this chapter, the silica nanosphere is levitated by a single optical tweezer beam propagating along the $z$ axis, while the optical cavity mode for pump-probe readout is aligned along the $x$ axis, as shown in Fig.~\ref{fig:5.1}. The trapping beam provides the primary confinement potential for the nanosphere, whereas the cavity field is significantly weaker and is utilized primarily for reading out the mechanical motion. Therefore, the transverse mechanical oscillations along the $x$ direction are predominantly determined by the optical potential generated by the trapping beam.

For a dielectric nanosphere situated in the Rayleigh regime, the optical dipole potential induced by the intensity distribution $I(\mathbf{r})$ is given by:
\begin{equation}
U(\mathbf{r}) = -\frac{\alpha}{2\epsilon_0 c}\,I(\mathbf{r}),
\end{equation}
where $\alpha$ is the polarizability of the sphere and $c$ is the speed of light. For a dielectric nanosphere of radius $r_s$ and relative permittivity $\epsilon$, the polarizability is defined as~\cite{doi:10.1073/pnas.0912969107}:
\begin{equation}
\alpha = 4\pi\epsilon_0 r_s^3 \frac{\epsilon-1}{\epsilon+2}.
\end{equation}

The optical tweezers can be modeled as a focused Gaussian beam propagating along the $z$ axis, with its intensity distribution near the focal spot expressed as:
\begin{equation}
I(x,y,z) = I_0 \frac{w_0^2}{w(z)^2} \exp\!\left[ -\frac{2(x^2+y^2)}{w(z)^2} \right],
\end{equation}
where $w_0$ is the beam waist and $I_0$ is the peak intensity at the focal center. Near the trap center $(x,y,z)=(0,0,0)$, and assuming the transverse displacement satisfies $x \ll w_0$, the intensity can be expanded as:
\begin{equation}
I(x) \approx I_0 \left( 1 - \frac{2x^2}{w_0^2} \right).
\end{equation}

Substituting this expression into the dipole potential and expanding with respect to $x$ up to the second order yields:
\begin{equation}
U(x) \approx U_0 + \frac{\alpha I_0}{\epsilon_0 c\,w_0^2}x^2,
\end{equation}
which corresponds to a harmonic oscillator potential:
\begin{equation}
U(x) = U_0 + \frac{1}{2}m_s\omega_x^2 x^2 ,
\end{equation}
where $m_s$ is the mass of the nanosphere. Comparing the coefficients directly yields the transverse mechanical frequency along the $x$ direction:
\begin{equation}
\omega_x = \sqrt{\frac{4\alpha I_0}{m_s\epsilon_0 c\,w_0^2}}.
\end{equation}
Utilizing the mass expression of the sphere $m_s = \frac{4}{3}\pi r_s^3\rho$ (where $\rho$ is the material density), the above outcome simplifies to:
\begin{equation}
\omega_x = \left( \frac{12 I_0}{\rho c\,w_0^2} \operatorname{Re}\frac{\epsilon-1}{\epsilon+2} \right)^{1/2}.
\end{equation}
The peak intensity $I_0$ of the trapping beam can be related to the optical power $P_{\mathrm{trap}}$ via the standard Gaussian beam relationship:
\begin{equation}
I_0 = \frac{2P_{\mathrm{trap}}}{\pi w_0^2}.
\end{equation}
Substituting this relation into the frequency formula yields a more convenient expression for evaluating the transverse mechanical frequency:
\begin{equation}
\omega_x = \left( \frac{24P_{\mathrm{trap}}}{\pi\rho c\,w_0^4} \operatorname{Re}\frac{\epsilon-1}{\epsilon+2} \right)^{1/2}.
\end{equation}

The cavity pump and probe fields propagating along the $x$ direction are orders of magnitude weaker than the trapping beam, and thus only introduce a minute perturbative correction to the trapping potential. To the lowest-order approximation, the transverse mechanical oscillation frequency is predominantly determined by the optical tweezers and is governed by the expression derived above~\cite{Gieseler_2013}. The numerical evaluation yields:
\begin{equation}
\omega_m = 2\pi \times 1.91 \times 10^{5}\,\mathrm{Hz}.
\end{equation}

With the transverse mechanical frequency established, the single-photon optomechanical coupling strength between the cavity field and the mechanical motion can be further estimated. Under the Rayleigh approximation, the presence of the dielectric nanosphere alters the effective refractive index of the cavity mode, leading to a small shift in the cavity resonance frequency. For a nanosphere of volume $V_s$, the modulation amplitude exerted on the cavity frequency can be written as:
\begin{equation}
g_{\mathrm{C}} = \frac{3V_s}{4V_c} \frac{\epsilon-1}{\epsilon+2} \,\omega_c ,
\end{equation}
where $V_c$ is the cavity mode volume. Because the cavity field forms a standing-wave distribution along the $x$ direction, the cavity resonance frequency as a function of the nanosphere position can be expressed as:
\begin{equation}
\omega_c(x) = \omega_c + g_{\mathrm{C}}\cos(2kx),
\end{equation}
where $k = 2\pi/\lambda$ is the cavity optical wavevector. When the nanosphere is trapped near an equilibrium position of the standing-wave field, this expression can be linearly expanded around the equilibrium point, yielding the derivative of the cavity frequency with respect to displacement:
\begin{equation}
\frac{\partial\omega_c}{\partial x} = 2k\,g_{\mathrm{C}} .
\end{equation}

On the other hand, the mechanical displacement operator can be expressed as:
\begin{equation}
x = x_{\mathrm{zpf}}(b+b^\dagger),
\end{equation}
where
\begin{equation}
x_{\mathrm{zpf}} = \sqrt{\frac{\hbar}{2m_s\omega_x}}
\end{equation}
represents the zero-point fluctuation amplitude of the mechanical resonator. Substituting this into the cavity frequency modulation term and comparing it with the standard optomechanical interaction Hamiltonian,
\begin{equation}
H_{\mathrm{int}} = \hbar g_0\,a^\dagger a (b+b^\dagger),
\end{equation}
one obtains the single-photon optomechanical coupling strength:
\begin{equation}
g_0 = 2k\,x_{\mathrm{zpf}}\, g_{\mathrm{C}} ,
\end{equation}
where
\begin{equation}
g_{\mathrm{C}} = \frac{3}{4}\frac{V_s}{V_c}\frac{\epsilon_r-1}{\epsilon_r+2}\,\omega_c = 1.28 \times 10^{4}\,\mathrm{Hz}.
\end{equation}
On this basis, the single-photon optomechanical coupling strength evaluates to:
\begin{equation}
g_0 = 3.70\,\mathrm{Hz}.
\end{equation}

The environmental damping is treated using the free-molecular regime approximation~\cite{doi:10.1073/pnas.0912969107,Hunger_2010}. In the numerical simulation, the residual gas pressure is taken as $P_{\mathrm{gas}} = 0.5 \times 10^{-7}\,\mathrm{Pa}$, the ambient temperature is set to $T = 150\,\mathrm{K}$, and the mass of the gas molecules is $m_{\mathrm{gas}} = 4.81 \times 10^{-26}\,\mathrm{kg}$. The corresponding mean thermal velocity is given by:
\begin{equation}
v = \sqrt{\frac{8k_B T}{\pi m_{\mathrm{gas}}}} = 3.31 \times 10^{2}\,\mathrm{m/s},
\end{equation}
from which the mechanical damping rate can be expressed as:
\begin{equation}
\gamma_m = \frac{3P_{\mathrm{gas}}}{r_s \rho v} = 2\pi \times 1.03 \times 10^{-5}\,\mathrm{Hz}.
\end{equation}

Based on the parameters established above, the Rabi frequency $\Omega_\text{pu}$ required for the transmission spectrum calculation can be determined as~\cite{9452539}:
\begin{equation}
\Omega_\text{pu} = \sqrt{\frac{2P_\text{pu}\kappa}{\hbar\omega_\text{pu}}} = 1.55 \times 10^{13}\,\mathrm{Hz}.
\end{equation}

\begin{figure}[ht!]
\centering
\includegraphics[width=0.7\textwidth]{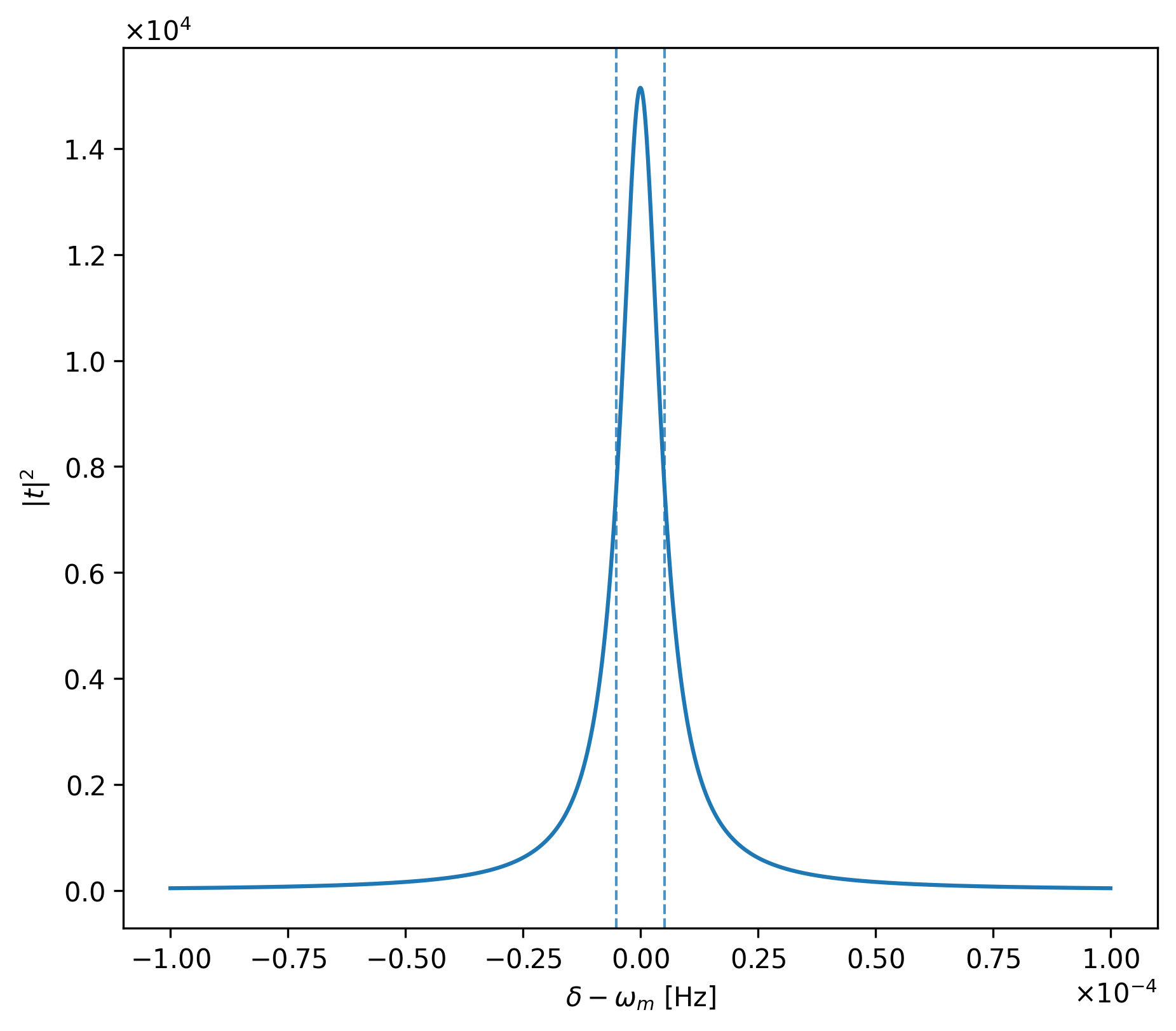}
\caption{The transmission of the probe field $|t|^2$ as a function of $\delta - \omega_m$, where the spectrum is centered at $\delta - \omega_m$. The dashed lines indicate the full width at half maximum (FWHM), which characterizes the minimum resolvable frequency difference of the resonance peak, given by $\omega_{\mathrm{FWHM}} \simeq 1.03 \times 10^{-5}\,\mathrm{Hz}$.}
\label{fig:5.3}
\end{figure}

In the subsequent calculation of the transmission spectrum, we further set the cavity-pump detuning to $\Delta = 0$. On this basis, the transmission spectrum is obtained as shown in Fig.~\ref{fig:5.3}, where the resonance peak appears at $\delta - \omega_m = 0$. The full width at half maximum (FWHM) can be expressed as:
\begin{equation}
\omega_{\mathrm{FWHM}} \simeq 1.03 \times 10^{-5} \, \mathrm{Hz} = \delta \omega_\text{min}.
\end{equation}
This width characterizes the resolution of the spectral line to frequency variations under the current parameter set, representing the minimum distinguishable frequency difference $\delta \omega_\text{min}$ that the resonance peak can resolve. It also serves as a benchmark for subsequently determining whether the frequency shift induced by the additional interaction is resolvable.

Based on this, we further investigate the response of the transmission spectrum when the mechanical frequency undergoes a minute variation. The fundamental underlying concept is that when the external force varies slowly with position—and thus can be approximated as locally linear within the motion range of the nanosphere—the motion of the levitated nanosphere within the trapping potential can be modeled as a simple harmonic oscillation. The intrinsic spring constant is denoted as $k$, corresponding to the mechanical resonance frequency:
\begin{equation}
\omega_n = \sqrt{\frac{k}{m}},
\end{equation}
where $m$ is the mass of the nanosphere.

In the presence of a non-zero force gradient within the system, this additional interaction modifies the equivalent restoring force experienced by the nanosphere, thereby correcting the effective spring constant to:
\begin{equation}
k' = k + \frac{\partial F}{\partial x}.
\end{equation}
Consequently, the corrected resonance frequency can be written as:
\begin{equation}
\omega_n' = \sqrt{\frac{k'}{m}} = \sqrt{\frac{k + \partial F/\partial x}{m}} = \omega_n \sqrt{1 + \frac{1}{k} \frac{\partial F}{\partial x}}.
\end{equation}

Under the condition that the frequency shift induced by the force gradient is sufficiently small, i.e., $\omega_n' \approx \omega_n$, the above expression can be expanded to first order, yielding:
\begin{equation}
\omega_n' \approx \omega_n \left( 1 + \frac{1}{2k} \frac{\partial F}{\partial x} \right).
\end{equation}
This leads to:
\begin{equation}
\frac{1}{2k} \frac{\partial F}{\partial x} \approx \frac{\omega_n'}{\omega_n} - 1.
\end{equation}

If we further define the frequency shift as $\Delta\omega = \omega_n' - \omega_n$, the approximate relationship between the force gradient and the frequency shift satisfies:
\begin{equation}
\frac{\partial F}{\partial x} = \frac{2k \, \Delta\omega}{\omega_n}.
\label{eq:do}
\end{equation}

\begin{figure}[ht!]
\centering
\includegraphics[width=0.7\textwidth]{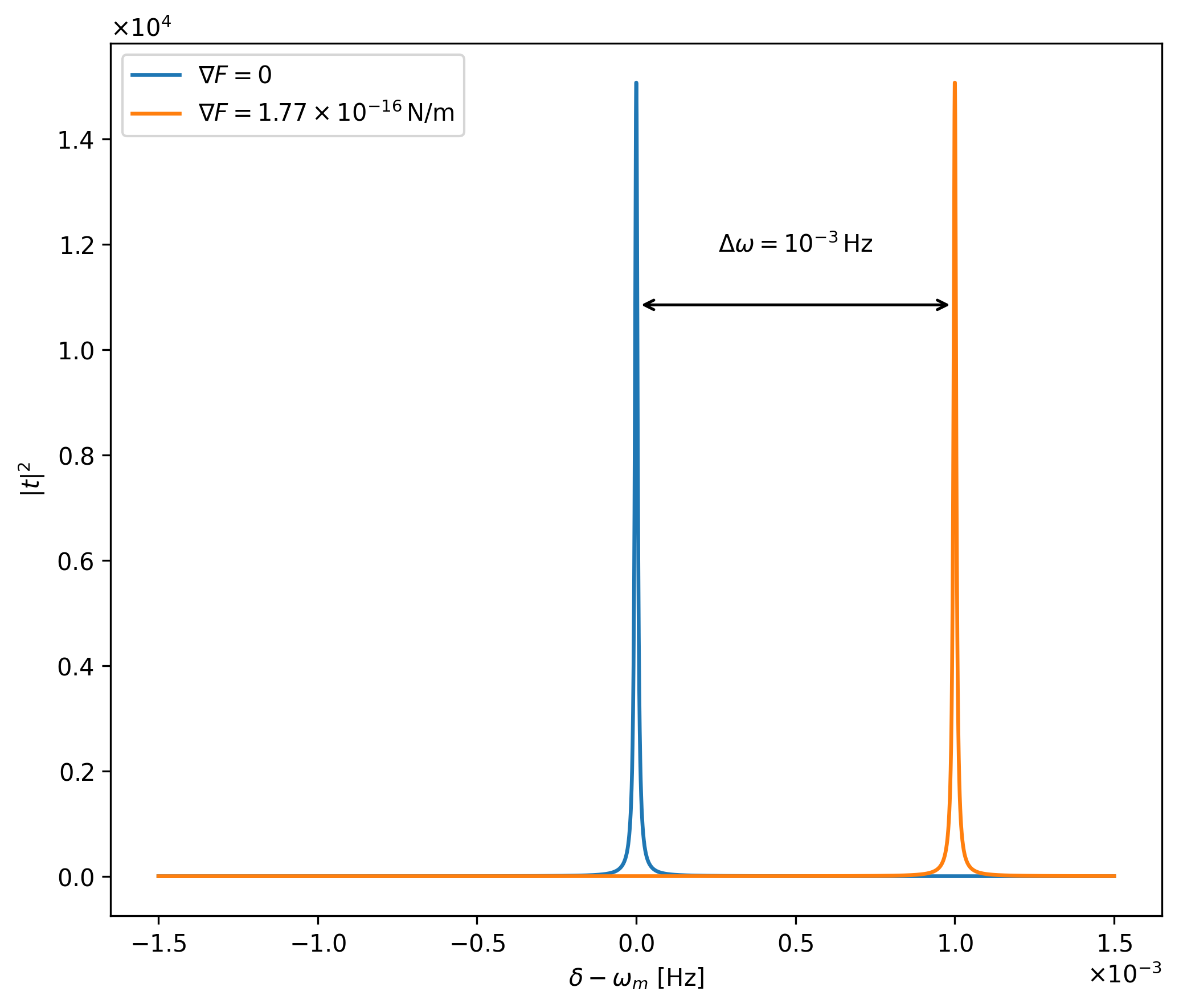}
\caption{Transmission $|t|^2$ of the probe field as a function of $\delta-\omega_m$. A resonance peak is observed at $\delta-\omega_m=0$ in the absence of external force gradient. When a force gradient $\nabla F = 1.77\times10^{-16}\,\mathrm{N/m}$ is applied, the resonance peak shifts to a finite frequency, demonstrating the measurable frequency shift induced by the external interaction.}
\label{fig:5.4}
\end{figure}

Taking the state with $\nabla F = 1.77 \times 10^{-16} \, \mathrm{N/m}$ as an example, the resonance peak shift induced by this force gradient is $\Delta \omega = 10^{-3} \, \mathrm{Hz}$. The corresponding results are illustrated in Fig.~\ref{fig:5.4}, where the original resonance peak remains near $\delta - \omega_m = 0$, whereas the frequency-shifted resonance peak moves to $\delta - \omega_m = 10^{-3} \, \mathrm{Hz}$. Consequently, a distinguishable double-peak structure can be formed by superimposing the photoelectric signals from the two transmission paths. In our proposed scheme, once a mechanical frequency correction due to the additional interaction occurs, the corresponding signal can be extracted by comparing the peak position difference between the reference and the perturbed spectral lines. Under the parameters chosen in this chapter, $\Delta \omega = 10^{-3} \, \mathrm{Hz}$ is significantly larger than the FWHM of a single-peak spectrum; thus, the two states can be clearly distinguished in the frequency domain. Our system translates the minute mechanical frequency variations induced by weak interactions into observable peak shifts, which subsequently serve as the readout for constraining the additional force or force gradient.

\subsection{Casimir force background and environmental noise analysis}\label{3.2}
In short-range precision measurements, the Casimir force is generally treated as a standard background interaction that must be modeled and subtracted, rather than a new physics signal to be constrained. In a sphere-plane system, the method for evaluating the Casimir force depends on the relative scales of the geometric parameters. The most common short-distance approximation is the proximity force approximation (PFA), the core idea of which is to treat the spherical surface as a collection of local parallel planar elements and integrate the plane-plane results along the local separation. For an ideal conductor sphere-plane geometry, the force given by the PFA satisfies
\begin{equation}
F_{\mathrm{PFA}}(z)\propto \frac{r_s}{z^3},
\end{equation}
where \(r_s\) is the radius of the sphere. Furthermore, under the condition \(z/r_s\ll 1\), curvature corrections can be incorporated on top of the PFA, which can be expressed, for instance, as
\begin{equation}
E(z)=E_{\mathrm{PFA}}(z)
\left[
1+\theta_1\frac{z}{r_s}
+\mathcal{O}\!\left(\frac{z^2}{r_s^2}\right)
\right].
\end{equation}
This correction remains inherently a short-distance expansion around the small parameter \(z/r_s\); thus, its applicability is still predicated on the assumption that
\begin{equation}
\frac{z}{r_s}\ll 1.
\end{equation}
If the system operates in the regime where \(r_s\ll z\), the sphere behaves more like a polarizable small scatterer relative to the plate, rendering the image of local parallel planar elements no longer appropriate. Under such circumstances, neither the PFA nor its curvature corrections provide the most natural approximation.

When the condition $r_s/z\ll 1$ holds, a more reasonable approach is to employ the Casimir--Polder approximation in the small-sphere limit. The underlying physical picture is that an induced dipole moment is generated in the small sphere under the combined action of vacuum fluctuations and the reflected fields, which subsequently interacts with the electromagnetic fields modified by the plate. Here, the primary material information of the small sphere is encapsulated in its dynamic polarizability along the imaginary frequency axis:
\begin{equation}
\alpha(i\xi)
=
4\pi\varepsilon_0 r_s^3
\frac{\varepsilon_s(i\xi)-1}{\varepsilon_s(i\xi)+2},
\end{equation}
where \(\varepsilon_s(i\xi)\) denotes the dielectric function of the sphere material on the imaginary frequency axis. For the plate, the material response is characterized by the Fresnel reflection coefficients. Defining the magnitude of the wave vector parallel to the plate surface as \(k\), and introducing
\begin{equation}
\kappa=\sqrt{k^2+\frac{\xi^2}{c^2}},
\qquad
\kappa_p=
\sqrt{
k^2+\varepsilon_p(i\xi)\mu_p(i\xi)\frac{\xi^2}{c^2}
},
\end{equation}
the reflection coefficients corresponding to the TE and TM polarizations can be written as
\begin{equation}
r_{\mathrm{TE}}(i\xi,k)
=
\frac{\mu_p(i\xi)\kappa-\kappa_p}
{\mu_p(i\xi)\kappa+\kappa_p},
\end{equation}
\begin{equation}
r_{\mathrm{TM}}(i\xi,k)
=
\frac{\varepsilon_p(i\xi)\kappa-\kappa_p}
{\varepsilon_p(i\xi)\kappa+\kappa_p},
\end{equation}
respectively, where \(\varepsilon_p(i\xi)\) and \(\mu_p(i\xi)\) are the dielectric function and magnetic permeability of the plate material on the imaginary frequency axis. For typical non-magnetic media, one can set \(\mu_p(i\xi)=1\).

Under the aforementioned approximations, the Casimir potential energy of the sphere-plane system can be expressed as
\begin{equation}
U(z)
=
\frac{\hbar\mu_0}{8\pi^2}
\int_0^\infty d\xi\,\xi^2\,\alpha(i\xi)
\int_0^\infty dk\,\frac{k}{\kappa}\,e^{-2\kappa z}
\left[
r_{\mathrm{TE}}(i\xi,k)
-
\left(
1+2\frac{c^2k^2}{\xi^2}
\right)
r_{\mathrm{TM}}(i\xi,k)
\right].
\end{equation}
The resulting Casimir force is given by
\begin{equation}
F_C(z)=-\frac{dU(z)}{dz},
\end{equation}
which yields
\begin{equation}
F_C(z)
=
\frac{\hbar\mu_0}{4\pi^2}
\int_0^\infty d\xi\,\xi^2\,\alpha(i\xi)
\int_0^\infty dk\,k\,e^{-2\kappa z}
\left[
r_{\mathrm{TE}}(i\xi,k)
-
\left(
1+2\frac{c^2k^2}{\xi^2}
\right)
r_{\mathrm{TM}}(i\xi,k)
\right].
\end{equation}
If one is further interested in the force gradient, differentiating with respect to the distance yields
\begin{equation}
\frac{\partial F_C}{\partial z}
=
-\frac{\hbar\mu_0}{2\pi^2}
\int_0^\infty d\xi\,\xi^2\,\alpha(i\xi)
\int_0^\infty dk\,k\kappa\,e^{-2\kappa z}
\left[
r_{\mathrm{TE}}(i\xi,k)
-
\left(
1+2\frac{c^2k^2}{\xi^2}
\right)
r_{\mathrm{TM}}(i\xi,k)
\right].
\end{equation}
To evaluate the dominant contribution to the interaction, the following approximations can be adopted:
\begin{equation}
\alpha(i\xi)\approx \alpha(0),
\qquad
\varepsilon_p(i\xi)\approx \varepsilon_p(0).
\end{equation}
For an isotropic, non-magnetic dielectric sphere, its zero-frequency polarizability can be further simplified to
\begin{equation}
\alpha(0)
=
4\pi\varepsilon_0 r_s^3
\frac{\varepsilon_s(0)-1}{\varepsilon_s(0)+2}.
\end{equation}
Assuming an ideal conducting plate as a further approximation, the Fresnel reflection coefficients become
\begin{equation}
r_{\mathrm{TE}}=-1,
\qquad
r_{\mathrm{TM}}=1.
\end{equation}
Consequently, the potential energy integral can be evaluated analytically, yielding the classic Casimir--Polder results:
\begin{equation}
U(z)=
-\frac{3\hbar c}{8\pi}\,
\frac{\varepsilon_s(0)-1}{\varepsilon_s(0)+2}\,
\frac{r_s^3}{z^4},
\end{equation}
and
\begin{equation}
F_C(z)=
-\frac{3\hbar c}{2\pi}\,
\frac{\varepsilon_s(0)-1}{\varepsilon_s(0)+2}\,
\frac{r_s^3}{z^5}.
\end{equation}
Based on these analytical results, the frequency shift induced by the gradient of the Casimir force can be subtracted as a background.

Once the spectral line position is established as a readout, it is necessary to evaluate the resolvability of this frequency shift against the noise background. For current optically levitated nanosphere systems, the frequency tracking of the mechanical mode is ultimately limited by thermal fluctuations. Therefore, it is essential to incorporate the minimum resolvable frequency shift corresponding to thermal noise into the parametric model \cite{10.1143/PTP.49.1516,10.1063/1.1642738}. To estimate this minimum resolvable frequency shift, one must start from the thermal fluctuation spectrum of the mechanical oscillator, convert it into a frequency noise spectrum, and evaluate the corresponding mean-square fluctuations within a finite measurement bandwidth.

For a one-dimensional mechanical oscillator, the equation of motion is given by
\begin{equation}
m_{\mathrm{eff}}\ddot{x}
+m_{\mathrm{eff}}\frac{\omega_n}{Q}\dot{x}
+m_{\mathrm{eff}}\omega_n^2 x
=
F_{\mathrm{th}}(t),
\end{equation}
where $m_{\mathrm{eff}}$ is the effective mass, $\omega_n$ is the intrinsic mechanical angular frequency, $Q$ is the mechanical quality factor, and $F_{\mathrm{th}}(t)$ represents the thermal noise force. Transforming this equation into the frequency domain yields
\begin{equation}
x(\omega)=\chi(\omega)F_{\mathrm{th}}(\omega),
\end{equation}
where the mechanical response function is defined as
\begin{equation}
\chi(\omega)=\frac{1}{m_{\mathrm{eff}}\left(\omega_n^2-\omega^2-i\omega\omega_n/Q\right)}.
\end{equation}
Accordingly, the displacement power spectral density can be expressed as
\begin{equation}
S_x(\omega)=|\chi(\omega)|^2S_F(\omega)
=
\frac{S_F(\omega)}
{m_{\mathrm{eff}}^2\left[(\omega^2-\omega_n^2)^2+\omega^2\omega_n^2/Q^2\right]}.
\label{eq:Sx_append}
\end{equation}

The thermal noise force spectrum satisfies the fluctuation-dissipation theorem \cite{10.1063/1.1499745,Robins1984PhaseNI}:
\begin{equation}
S_F(\omega)=4m_{\mathrm{eff}}\Gamma k_B T,
\end{equation}
where $\Gamma=\omega_n/Q$ is the mechanical damping rate, which leads to
\begin{equation}
S_F(\omega)=\frac{4m_{\mathrm{eff}}\omega_n k_B T}{Q}.
\label{eq:SF_append}
\end{equation}

To relate the displacement noise to the resonance frequency fluctuations, we further define the frequency noise spectral density as
\begin{equation}
S_{\omega}(\omega)
=
\left(\frac{\omega_n}{2Q}\right)^2
\frac{S_x(\omega)}{\langle x_{\mathrm{rms}}\rangle^2},
\label{eq:Somega_append}
\end{equation}
where $\langle x_{\mathrm{rms}}\rangle$ is the root-mean-square (RMS) amplitude of the oscillator. Near resonance, the dominant contribution originates from the frequency range around $\omega\simeq\omega_n$. Setting $\omega=\omega_n$, Eq.~\eqref{eq:Sx_append} yields
\begin{equation}
S_x(\omega_n)
=
\frac{S_F(\omega_n)}
{m_{\mathrm{eff}}^2\omega_n^4/Q^2}.
\end{equation}
Combining this with Eq.~\eqref{eq:SF_append} gives
\begin{equation}
S_x(\omega_n)
=
\frac{4k_B T\,Q}{m_{\mathrm{eff}}\omega_n^3}.
\end{equation}
Substituting this result into Eq.~\eqref{eq:Somega_append} yields the frequency noise spectrum near the resonance point:
\begin{equation}
S_{\omega}(\omega_n)
=
\frac{k_B T}
{m_{\mathrm{eff}}\omega_n \langle x_{\mathrm{rms}}\rangle^2 Q}.
\label{eq:Somega_res_append}
\end{equation}

Assuming a measurement bandwidth of $\Delta f$ within which $S_{\omega}(\omega)$ varies minimally, the mean-square value of the frequency fluctuations can be approximated as
\begin{equation}
(\Delta\omega_n)^2
\approx
S_{\omega}(\omega_n)\Delta f.
\end{equation}
Consequently, the minimum resolvable angular frequency shift can be expressed as
\begin{equation}
\Delta\omega_n
\approx
\sqrt{
\frac{k_B T\,\Delta f}
{m_{\mathrm{eff}}\omega_n \langle x_{\mathrm{rms}}\rangle^2 Q}
}.
\label{eq:domega_append}
\end{equation}

The corresponding mechanical quality factor is given by
\begin{equation}
Q=\frac{\omega_m}{\gamma_m}\approx1.85\times10^{10}.
\end{equation}
The RMS amplitude of the oscillator satisfies the relation $\langle x_{\mathrm{rms}}\rangle^2 < w_c^2/2$, and we set $\langle x_{\mathrm{rms}}\rangle \sim 0.5\,\mu\mathrm{m}$ here. Meanwhile, the measurable bandwidth $\Delta f\approx10^{-5}\,\mathrm{Hz}$ is determined by the characteristic response time of the oscillator $\tau\approx1.6\times10^{4}\,\mathrm{s}$ via the relation $\Delta f \approx 1/(2\pi \tau)$ \cite{articlejl}. Substituting these parameters yields the lower bound of the frequency resolution limited by thermal noise as $\delta \omega_{\mathrm{th}}\sim 3\times10^{-6}\,\mathrm{Hz}$.

\begin{figure}[ht!]
\centering\includegraphics[width=0.7\textwidth]{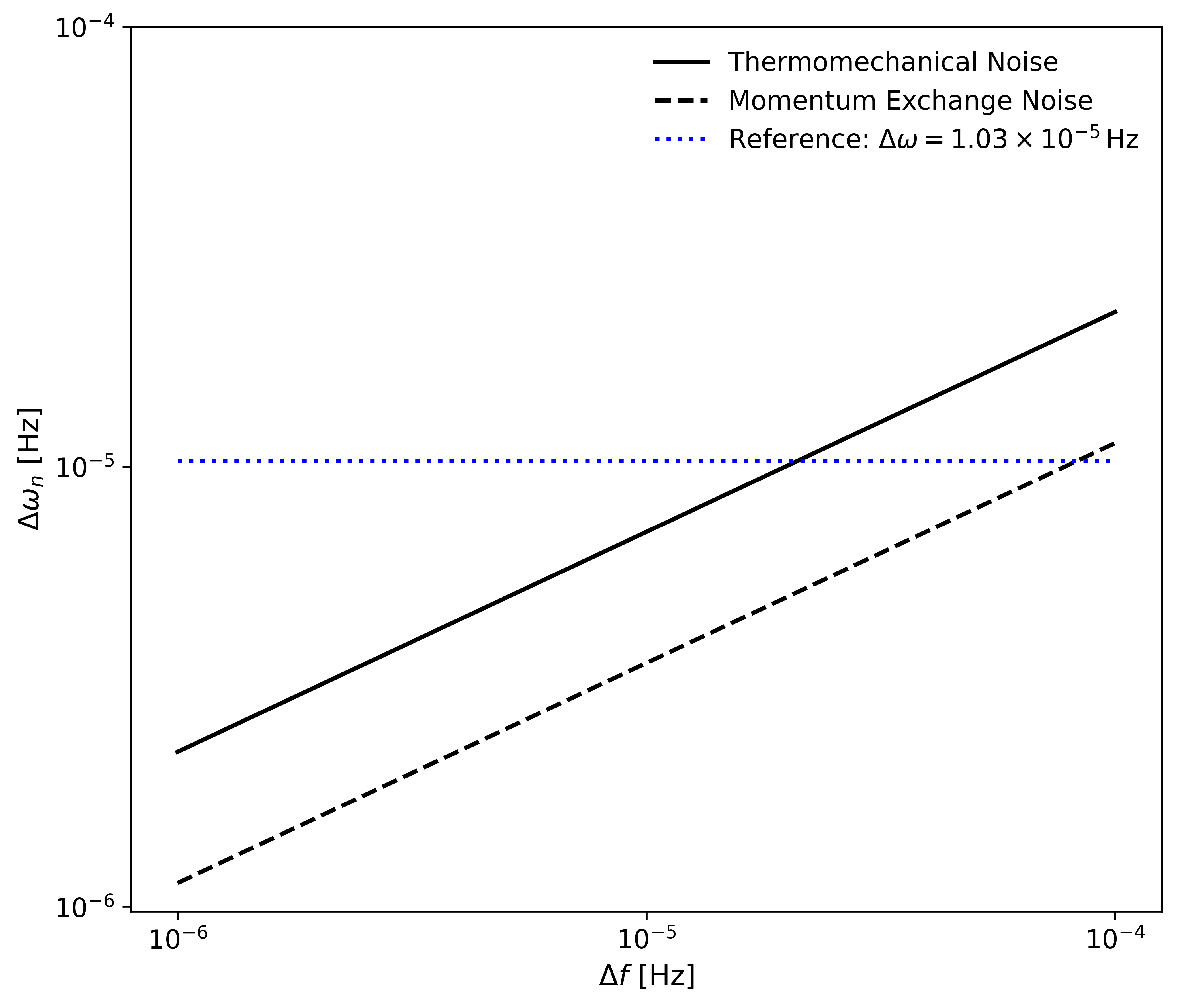}
\caption{Sensitivity limits due to thermomechanical fluctuations and momentum exchange noise compared to the detection limit from FWHM. With the main noise limits staying below the detection result, we can ignore their influence on the system safely.}
\label{fig:a1}
\end{figure}

Here, Fig.~\ref{fig:a1} illustrates a comparison of the thermal noise and momentum exchange noise against the previously derived detection limit. The momentum exchange noise arises from the interaction between residual gas molecules and the harmonic oscillator \cite{10.1063/1.1642738}. The calculation of the frequency shift $\Delta \omega$ induced by momentum exchange noise follows a procedure essentially identical to Eq.~(\ref{eq:domega_append}), except that the quality factor considering gas dissipation is defined as $Q_\text{gas} = m_\text{eff} \omega_n v / pA$, where $v = \sqrt{k_B T / m}$ is the thermal velocity of the gas molecules, $p$ is the gas pressure, and $A = 4\pi r_s^2$ is the surface area of the nanosphere. Compared to the limit imposed by the transmission linewidth $\delta \omega_\text{min}\simeq 1.03\times10^{-5}\,\mathrm{Hz}$, the resolution floor restricted by the noise here is significantly lower. Consequently, under our choice of parameters, the impact of the primary noise sources can be safely neglected.

\subsection{Expected constraints}\label{3.3}
The core ideal of setting constraints is that if an additional interaction introduced by axions exists between the small sphere and the nearby source mass, the force gradient of this interaction will induce an intrinsic frequency shift in the mechanical oscillator. This relationship is given by Eq.~\ref{eq:do} mentioned above. If no anomalous frequency shift is observed, it indicates that the potential interaction falls below the lower limit of the experimental resolution, from which the upper bound for the additional interaction can be derived, namely
\begin{equation}
\Delta\left|
\frac{1}{2m_s\omega_0}\frac{\partial F_\text{add}(z)}{\partial z}
\right|
<
\delta \omega_\text{min}.
\end{equation}
This condition is based on the following rationale: it is assumed that the additional force gradient induced by a specific set of axion parameters is sufficiently large to shift the mechanical frequency beyond the experimental resolution. If no such signal is detected in the actual experiment, that parameter region is excluded.

More specifically, what is actually observed is not the force gradient from a single material, but the differential signal between the aluminum (Al) and silver (Ag) substrate mirrors:
\begin{equation}
\Delta\!\left(\frac{\partial F_{\mathrm{add}}}{\partial d}\right)
\equiv
\frac{\partial F^{\mathrm{Ag}}_{\mathrm{add}}}{\partial d}
-
\frac{\partial F^{\mathrm{Al}}_{\mathrm{add}}}{\partial d}.
\end{equation}
Under a simplified model with a uniform gold capping layer, the above expression can be written as
\begin{equation}
\Delta\!\left(\frac{\partial F_{\mathrm{add}}}{\partial d}\right)
=
\frac{2\pi r_s^3 m_a}{3m^2m_H^2}
\left(C_{\mathrm{Ag}}-C_{\mathrm{Al}}\right)C_{s}\,I(m_a).
\label{eq:diff_grad_general}
\end{equation}
For the SiO$_2$ sphere, the parameters in the effective material coefficients are given by \cite{PhysRevD.89.035010}:
\begin{equation}
\frac{Z_{s}}{\mu_{s}}=0.503205,\qquad
\frac{N_{s}}{\mu_{s}}=0.505179,\qquad
\rho_{s}=1.1\times10^{-5}\,(\mathrm{MeV})^{4}.
\end{equation}
For the Al substrate, we take:
\begin{equation}
\frac{Z_{\mathrm{Al}}}{\mu_{\mathrm{Al}}}=0.48558,\qquad
\frac{N_{\mathrm{Al}}}{\mu_{\mathrm{Al}}}=0.52304,\qquad
\rho_{\mathrm{Al}}=1.2\times10^{-5}\,(\mathrm{MeV})^{4}.
\end{equation}
For the Ag substrate, we take:
\begin{equation}
\frac{Z_{\mathrm{Ag}}}{\mu_{\mathrm{Ag}}}=0.439,\qquad
\frac{N_{\mathrm{Ag}}}{\mu_{\mathrm{Ag}}}=0.570,\qquad
\rho_{\mathrm{Ag}}=4.5\times10^{-5}\,(\mathrm{MeV})^{4}.
\end{equation}
The integral $I(m_a)$ in this model is defined as:
\begin{equation}
I(m_a)=
\int_{1}^{\infty}\!du\,
\frac{\sqrt{u^{2}-1}}{u}
\left(1-e^{-2m_a u D}\right)
e^{-2m_a u d},
\label{eq:I_ma}
\end{equation}
where the first term in Eq.~\eqref{eq:I_ma},
\begin{equation}
1-e^{-2m_a u D},
\end{equation}
reflects the depth of contribution from the substrate of finite thickness, while the second term,
\begin{equation}
e^{-2m_a u d},
\end{equation}
characterizes the distance dependence, where the effective distance is $d=a+t$, $a=0.2\,\mu\mathrm{m}$ is the distance from the sphere to the cavity mirror surface, $t=0.1\,\mu\mathrm{m}$ is the thickness of the gold capping layer on the mirror surface, and $D=1\,\mathrm{mm}$ is the thickness of the Al/Ag substrate mirror behind the coating. It should be noted that while a uniform gold coating ensures identical surface electromagnetic conditions for light propagation inside the cavity, it causes the coupled signal to attenuate as the coating thickness increases.

Under three common coupling hypotheses, Eq.~\eqref{eq:diff_grad_general} can be further cast into more explicit forms. When the proton coupling dominates, i.e.,
\begin{equation}
g_{ap}^{2}\gg g_{an}^{2},
\end{equation}
the differential force gradient can be expressed as
\begin{equation}
\Delta\!\left(\frac{\partial F_{\mathrm{add}}}{\partial d}\right)_{p}
=
\frac{2\pi r_s^3 m_a}{3m^2m_H^2}
\left(\frac{g_{ap}^{2}}{4\pi}\right)^{2}
\left(
\rho_{\mathrm{Ag}}\frac{Z_{\mathrm{Ag}}}{\mu_{\mathrm{Ag}}}
-
\rho_{\mathrm{Al}}\frac{Z_{\mathrm{Al}}}{\mu_{\mathrm{Al}}}
\right)
\left(
\rho_{s}\frac{Z_{s}}{\mu_{s}}
\right)
I(m_a).
\label{eq:diff_grad_p}
\end{equation}

When the neutron coupling dominates, i.e.,
\begin{equation}
g_{an}^{2}\gg g_{ap}^{2},
\end{equation}
the differential force gradient becomes
\begin{equation}
\Delta\!\left(\frac{\partial F_{\mathrm{add}}}{\partial d}\right)_{n}
=
\frac{2\pi r_s^3 m_a}{3m^2m_H^2}
\left(\frac{g_{an}^{2}}{4\pi}\right)^{2}
\left(
\rho_{\mathrm{Ag}}\frac{N_{\mathrm{Ag}}}{\mu_{\mathrm{Ag}}}
-
\rho_{\mathrm{Al}}\frac{N_{\mathrm{Al}}}{\mu_{\mathrm{Al}}}
\right)
\left(
\rho_{s}\frac{N_{s}}{\mu_{s}}
\right)
I(m_a).
\label{eq:diff_grad_n}
\end{equation}

When the proton and neutron couplings are symmetric, i.e.,
\begin{equation}
g_{ap}=g_{an}\equiv g_a,
\end{equation}
one obtains
\begin{equation}
\Delta\!\left(\frac{\partial F_{\mathrm{add}}}{\partial d}\right)_{\mathrm{eq}}
=
\frac{2\pi r_s^3 m_a}{3m^2m_H^2}
\left(\frac{g_{a}^{2}}{4\pi}\right)^{2}
\left[
\rho_{\mathrm{Ag}}
\left(
\frac{Z_{\mathrm{Ag}}}{\mu_{\mathrm{Ag}}}
+
\frac{N_{\mathrm{Ag}}}{\mu_{\mathrm{Ag}}}
\right)
-
\rho_{\mathrm{Al}}
\left(
\frac{Z_{\mathrm{Al}}}{\mu_{\mathrm{Al}}}
+
\frac{N_{\mathrm{Al}}}{\mu_{\mathrm{Al}}}
\right)
\right]
\left[
\rho_{s}
\left(
\frac{Z_{s}}{\mu_{s}}
+
\frac{N_{s}}{\mu_{s}}
\right)
\right]
I(m_a).
\label{eq:diff_grad_eq}
\end{equation}

In practical evaluations, for instance, in the proton-dominated scenario, from the condition
\begin{equation}
\left|
\Delta\!\left(\frac{\partial F_{\mathrm{add}}}{\partial d}\right)_{p}
\right|
<
2m_s\omega_0\,\delta\omega_{\min},
\end{equation}
one can solve for the coupling constant:
\begin{equation}
\frac{g_{ap}^{2}}{4\pi}
<
m\,m_H
\sqrt{
\frac{
3m_s\omega_0\,\delta\omega_{\min}
}{
\pi r_s^3 m_a
\left|
\left(
\rho_{\mathrm{Ag}}\frac{Z_{\mathrm{Ag}}}{\mu_{\mathrm{Ag}}}
-
\rho_{\mathrm{Al}}\frac{Z_{\mathrm{Al}}}{\mu_{\mathrm{Al}}}
\right)
\left(
\rho_{s}\frac{Z_{s}}{\mu_{s}}
\right)
I(m_a)
\right|
}
}.
\end{equation}

The neutron-dominated and symmetric coupling scenarios can be formulated in an entirely analogous manner, simply by replacing the material factors in the denominator with their corresponding forms from Eq.~\eqref{eq:diff_grad_n} and Eq.~\eqref{eq:diff_grad_eq}.

\begin{figure}[ht!]
\centering
\includegraphics[width=0.7\textwidth]{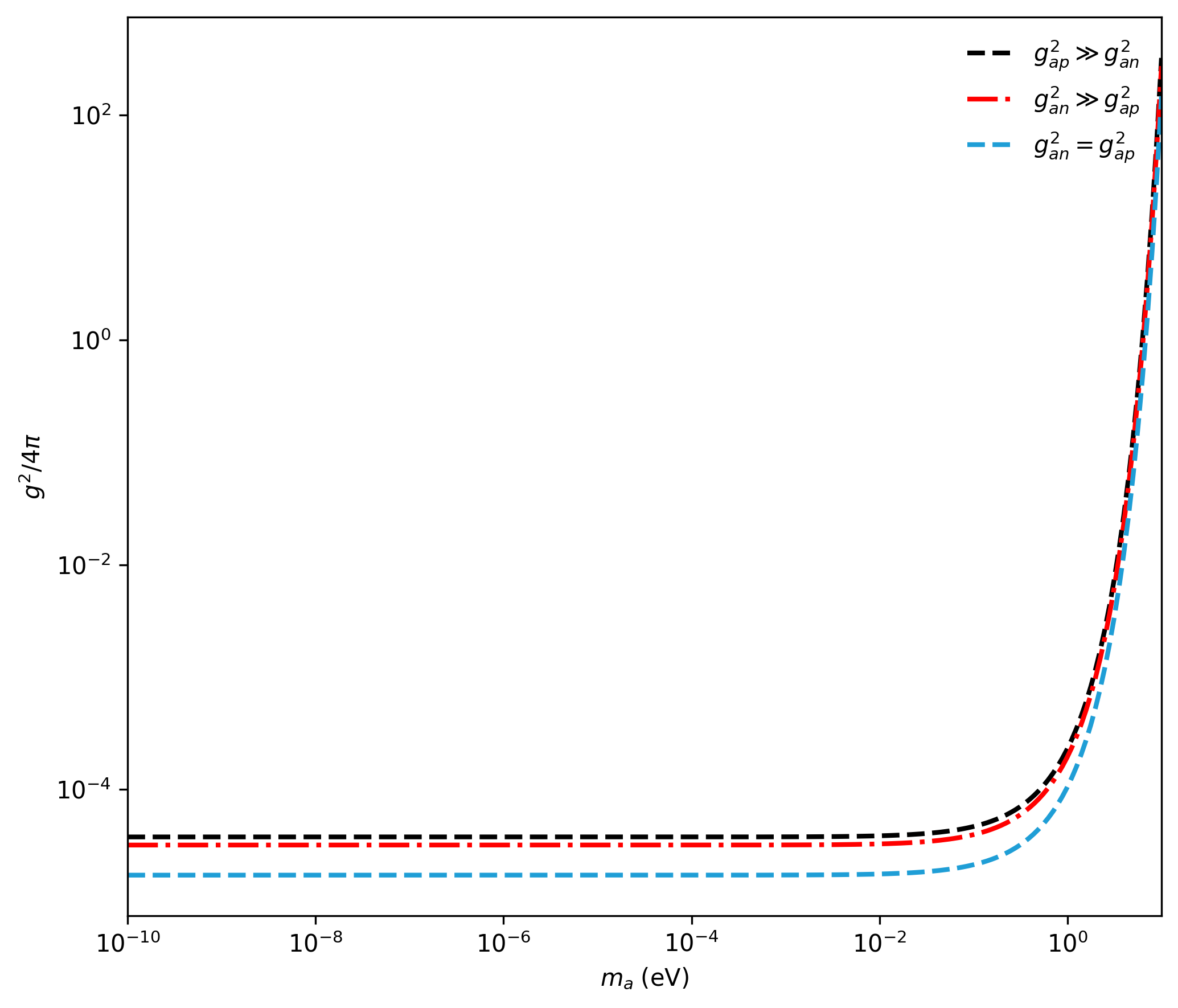}
\caption{Constraints on the axion–nucleon coupling constant as functions of the axion mass $m_a$ derived for the Ag/Al differential configuration with a common Au capping layer. The dashed black curve corresponds to the case $g_{ap}^{2} \gg g_{an}^{2}$, the dashed red curve to $g_{an}^{2} \gg g_{ap}^{2}$, and the blue dashed curve to the symmetric case $g_{an}^{2}=g_{ap}^{2}$}
\label{fig:5.5}
\end{figure}

The results derived from the aforementioned calculation method are presented in Fig.~\ref{fig:5.5}. The discrepancies among the three scenarios originate entirely from the different nucleon weightings within the material combinations. Given the chosen parameters for Ag, Al, and SiO$_2$, the differential force gradient induced by the Ag/Al material pair is typically largest in the symmetric coupling scenario, followed by the neutron-dominated case, and is relatively weakest in the proton-dominated scenario. This indicates that for the configuration utilizing aluminum and silver substrate mirrors, a more prominent differential force gradient signal can be achieved if the axion coupling to neutrons is at least as strong as its coupling to protons.

\begin{figure}[ht!]
\centering
\includegraphics[width=0.7\textwidth]{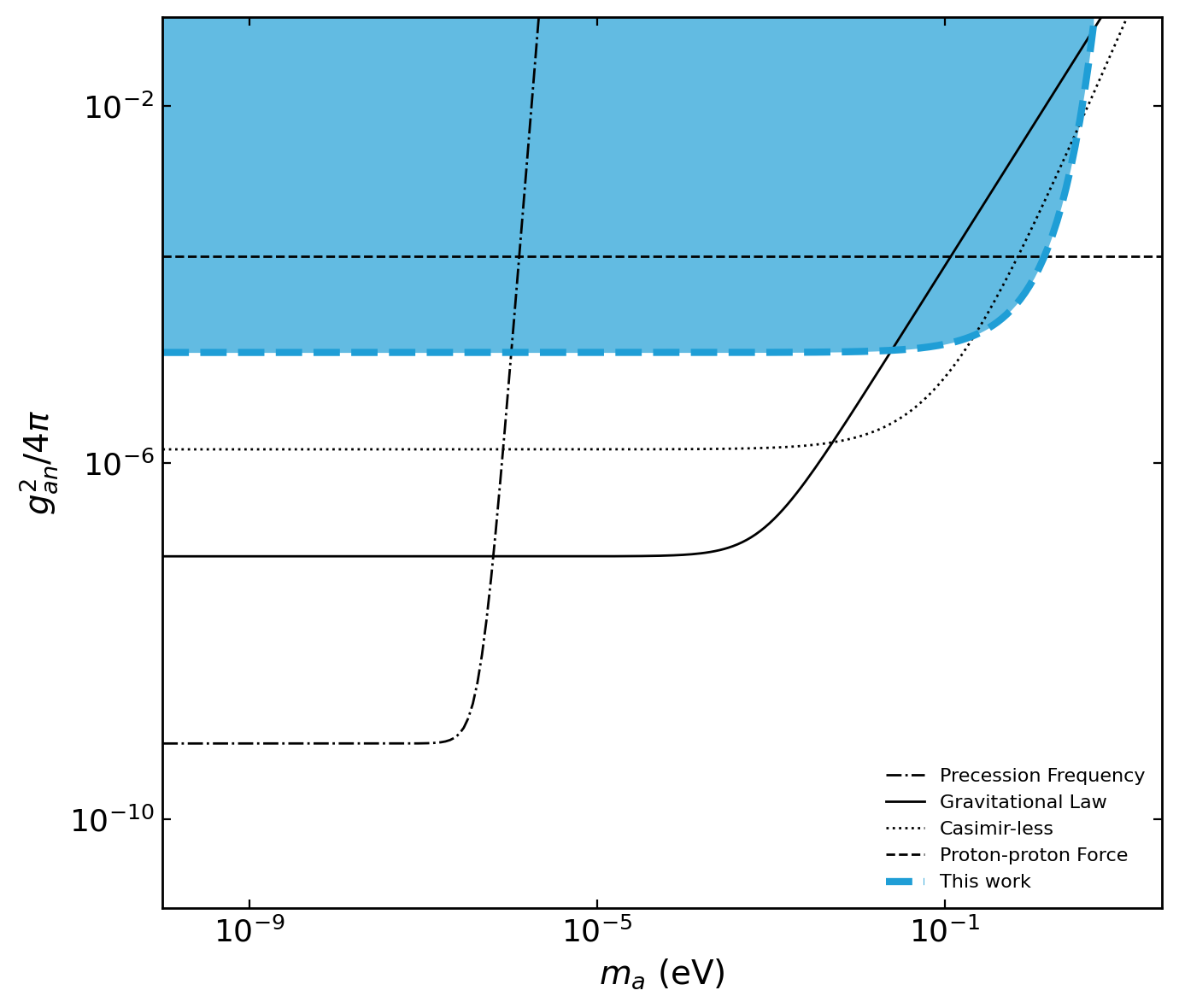}
\caption{Constraints on the axion–neutron coupling constant as a function of the axion mass $m_a$, assuming $g_{an}^{2}=g_{ap}^{2}$. The constraints here result from measurements of changes in the precession frequency \cite{PhysRevLett.103.261801}, tests of the gravitational inverse-square law \cite{PhysRevLett.98.131104,PhysRevLett.98.021101}, the Casimir-less experiment \cite{PhysRevLett.116.221102,Klimchitskaya2015} and the proton–proton force measurements \cite{RAMSEY1979285,PhysRevLett.110.040402}, respectively. The blue dashed curve shows the sensitivity obtained in this work. The blue shaded region indicates the excluded parameter space.}
\label{fig:5.6}
\end{figure}

Figure~\ref{fig:5.6} displays the constraints on the axion--neutron coupling constant as a function of the axion mass $m_a$. The various curves in the plot represent the sensitivities of established experiments alongside that of this work, including constraints derived from spin-precession frequency measurements, gravitational inverse-square law tests, Casimir-less experiments, and proton--proton force measurements \cite{PhysRevLett.103.261801,PhysRevLett.98.131104,PhysRevLett.98.021101,PhysRevLett.116.221102,Klimchitskaya2015,RAMSEY1979285,PhysRevLett.110.040402}. The blue dashed curve indicates the anticipated sensitivity achieved in this work, and the blue shaded region marks the parameter space that has been excluded. Compared with existing results, the constraints established in this work exhibit superior performance in the range of $m_a = 0.1\,\mathrm{eV}$ to $1\,\mathrm{eV}$. Our constraints offer a meaningful complement to the axion--neutron coupling parameter space and extend the exploratory reach of current experiments into relevant parameter regions.

\section{Conclusion}\label{4}
This work proposes an axion–nucleon coupling detection scheme based on the optical readout of cavity optomechanics. For the interaction between unpolarized macroscopic test bodies, this chapter considers the spin-independent effective potential induced by two-axion exchange between nucleons, which is integrated over a sphere-plane geometry to obtain the additional force gradient acting on a levitated nanosphere. By analytically establishing the mapping between the intrinsic mechanical resonance frequency and the external force gradient, this additional interaction can be converted into a minute shift in the mechanical resonance frequency. Furthermore, by incorporating the changes in the resonance peak position within the pump–probe transmission spectrum, the optical readout of this frequency shift signal can be achieved. On this basis, according to the system sensitivity to the minimum resolvable shift of the resonance peak position, this chapter evaluates the detectable additional force gradient under different source mass configurations, thereby yielding the anticipated constraints in the parameter space of axion mass and coupling constants. The results demonstrate that, taking the hypothesis $g_{an}^{2}=g_{ap}^{2}$ as an example, within the mass range of $m_a \approx 0.1\,\mathrm{eV}$ to $1\,\mathrm{eV}$, the scheme proposed in this chapter extends the current constraint boundaries by up to approximately two orders of magnitude. This indicates that the cavity optomechanical frequency-shift readout method possesses significant application potential in precision measurements of axion–nucleon interactions.

In terms of experimental realization and measurement sensitivity, there remains further room for optimization. For instance, employing an optical cavity with a larger mode volume and higher finesse, combined with cavity-assisted cooling techniques~\cite{PhysRevA.91.013824,PhysRevLett.118.183601,PhysRevD.98.030001,PhysRevA.64.033804}, is expected to yield a narrower spectral lineshape, thereby enhancing the resolution for frequency shifts. Regarding numerical evaluations, implementing more precise numerical relaxation methods~\cite{article2,PhysRevD.94.044051} and adopting appropriate approximation treatments for complex geometries~\cite{PhysRevD.76.124034,PhysRevLett.98.050403} will also contribute to improving the reliability of theoretical predictions and the stability of parameter constraints.

Furthermore, extending this class of detection schemes to a broader physical context is of equal significance. For example, the unique characteristics of levitated cavity optomechanical platforms in high-sensitivity force measurements render them promising candidates for addressing precision measurement problems associated with gravitational waves~\cite{PhysRevD.109.123023,PhysRevD.109.124007}. Notably, recent progress in the coherent manipulation of massive nanoparticles, such as matter-wave interferometry~\cite{Gerlich2026} and quantum squeezing in levitated optomechanical systems~\cite{doi:10.1126/science.ady4652}, offers the potential to further improve system coherence, increase the test mass, and suppress measurement noise. Benefiting from these technological developments, cavity-optomechanically based precision measurement schemes are expected to impose more stringent experimental constraints on various models in the future. Along with continuous advances in experimental techniques, we believe that more practical optomechanics-based detection devices will emerge in the near future.

\begin{backmatter}
\bmsection{Funding}
Natural Science Foundation of Shanghai (Grant No. 20ZR1429900).

\bmsection{Acknowledgments}
This work is supported by Natural Science Foundation of Shanghai (Grant No. 20ZR1429900).

\bmsection{Disclosures}
The authors declare no conflicts of interest.

\bmsection{Data Availability Statement}
The data that supports the findings of this study are available within the article.
\end{backmatter}

%%%%%%%%%%%%%%%%%%%%%%% References %%%%%%%%%%%%%%%%%%%%%%%%%

\bibliography{refoe}

%%%%%%%%%% If preparing manually:
% \begin{thebibliography}{1}
% \newcommand{\enquote}[1]{``#1''}

% \bibitem{Zhang:14}
% Y.~Zhang, S.~Qiao, L.~Sun, Q.~W. Shi, W.~Huang, L.~Li, and Z.~Yang,
%   \enquote{Photoinduced active terahertz metamaterials with nanostructured
%   vanadium dioxide film deposited by sol-gel method,}
%   {\protect\JournalTitle{Optics Express}} \textbf{22}, 11070--11078 (2014).

% \bibitem{Optica}
% {Optica}, \enquote{{Optica Publishing Group},}
%   \url{http://www.opg.optica.org}.

% \bibitem{FORSTER2007}
% P.~Forster, V.~Ramaswamy, P.~Artaxo, T.~Bernsten, R.~Betts, D.~Fahey,
%   J.~Haywood, J.~Lean, D.~Lowe, G.~Myhre, J.~Nganga, R.~Prinn, G.~Raga,
%   M.~Schulz, and R.~V. Dorland, \enquote{Changes in atmospheric consituents and
%   in radiative forcing,} in \enquote{Climate Change 2007: The Physical Science
%   Basis. Contribution of Working Group 1 to the Fourth assesment report of
%   Intergovernmental Panel on Climate Change,}  S.~Solomon, D.~Qin, M.~Manning,
%   Z.~Chen, M.~Marquis, K.~B. Averyt, M.~Tignor, and H.~L. Miler, eds.
%   (Cambridge University Press, 2007).

% \end{thebibliography}

\end{document}